\documentclass[letterpaper, 10 pt, conference]{ieeeconf}

\IEEEoverridecommandlockouts

\usepackage{graphics}
\usepackage{epsfig}
\usepackage{amsmath}
\usepackage{amssymb}

\usepackage{booktabs}

\usepackage{subcaption}
\newcommand{\SVC}{\mathrm{SVC}}

\newcommand{\LA}{\mathrm{LA}}

\let\labelindent\relax
\usepackage{enumitem}

\usepackage{cite}

\usepackage{multirow}

\title{\LARGE \bf
Motion-Sickness-Aware Vehicle Control \\
via DKUC-Based Inverse Design and Iterative Feedback Tuning }

\author{Uta Kawakami and Kenji Sawada
\thanks{This work was supported by JSPS KAKENHI Grant Numbers JP23H01636, JP23K03915, and JP22H01509, and JST CREST Grant Numbers JPMJCR23M4 and JPMJCR23M2.}
\thanks{U. Kawakami is with Graduate School of Engineering, The University of Osaka, 2-1 Yamadaoka, Suita, Osaka 565-0871, Japan {\tt\small uta.kawakami@misc.mech.eng.osaka-u.ac.jp}} \thanks{K. Sawada is with Graduate School of Engineering, The University of Osaka, 2-1 Yamadaoka, Suita, Osaka 565-0871, Japan {\tt\small knj.sawada@mech.eng.osaka-u.ac.jp}} }

\begin{document}

\maketitle
\thispagestyle{empty}
\pagestyle{empty}

\begin{abstract}
    Tuning existing vehicle controllers for motion-sickness reduction requires a desired vehicle response that reflects physiological comfort.
    The six-degree-of-freedom subjective vertical conflict (6DoF-SVC) model can evaluate sensory conflict induced by vehicle motion, but it does not directly provide a target motion for controller design.
    This paper proposes a 2-step framework to construct such a reference.
    First, a target SVC trajectory is generated from the baseline response by limiting conflict peaks while imposing a minimum activity level.
    Second, a lifted linear model identified using Deep KoopmanU with Control (DKUC) is inverted to map the target SVC trajectory to a regularized reference expressed in vehicle-acceleration and angular-velocity coordinates.
    The resulting motion reference is then used in an Iterative Feedback Tuning (IFT)-style procedure for fixed-structure PI and LQR controllers.
    In the considered simulation scenario, the selected SVC target reduces the sequence norm by 6.95\% relative to the baseline.
    Controller tuning with reduced response models decreases the regularized PI and LQR objectives by 49.5\% and 32.3\%, respectively.
    These results show that a nonlinear physiological comfort objective can be converted into a practical motion reference for tuning fixed-structure vehicle controllers.
\end{abstract}

\section{Introduction}
    Occupant comfort has become an important vehicle-control objective as automated driving allows occupants to read, work, or view displays.
    Low-frequency vehicle motion can cause motion sickness and reduce acceptance of automated vehicles \cite{zhang2024_its}.
    Conventional objectives based on weighted acceleration, vibration dose, or jerk are convenient for control design \cite{iso2631, sugiyama2025_iccae}, but do not explicitly model vestibular sensory conflict \cite{wada2025_bc, zhang2024_its}.
    The six-degree-of-freedom subjective vertical conflict (6DoF-SVC) model addresses this aspect by comparing sensed and internally estimated vertical directions \cite{bos1998_brb, buchheit2021_its}.

    Incorporating this physiological objective into an existing vehicle controller requires a link between two different quantities: sensory conflict and vehicle motion.
    The passenger--6DoF-SVC model predicts conflict from vehicle motion, whereas the controller acts through motor torques and steering.
    Thus, a desired conflict trajectory cannot be supplied directly as an actuator command.
    We address this gap by first converting the physiological target into a desired vehicle-motion response and then tuning the installed controller against that response.
    This allows the physiological objective to enter the design while retaining the installed speed- and path-tracking controller structures.

    Related approaches include nonlinear model predictive control (MPC) with SVC objectives \cite{todaka2024_nmpc}, ideal motion planning for tuning linear MPC weights \cite{todaka2025_arob}, and reinforcement-learning-based linear quadratic regulator (LQR) tuning with a coupled motion-sickness model \cite{yang2026_jauto}.
    Data-driven predictive control has also been studied for mixed vehicle platoons \cite{li2025_iotj}.
    Here, the focus is on constructing a motion reference while retaining existing proportional--integral (PI) and LQR controller structures.
    Iterative Feedback Tuning (IFT) and Fictitious Reference Iterative Tuning (FRIT) provide established ways to tune fixed controllers from response data \cite{hjalmarsson1998_cs, soma2004_frit}.
    Their use still requires a desired response or equivalent target, whose selection is difficult when the objective is specified by a nonlinear physiological model.

\begin{figure}[t]
        \centering
        \includegraphics[width=0.57\linewidth]{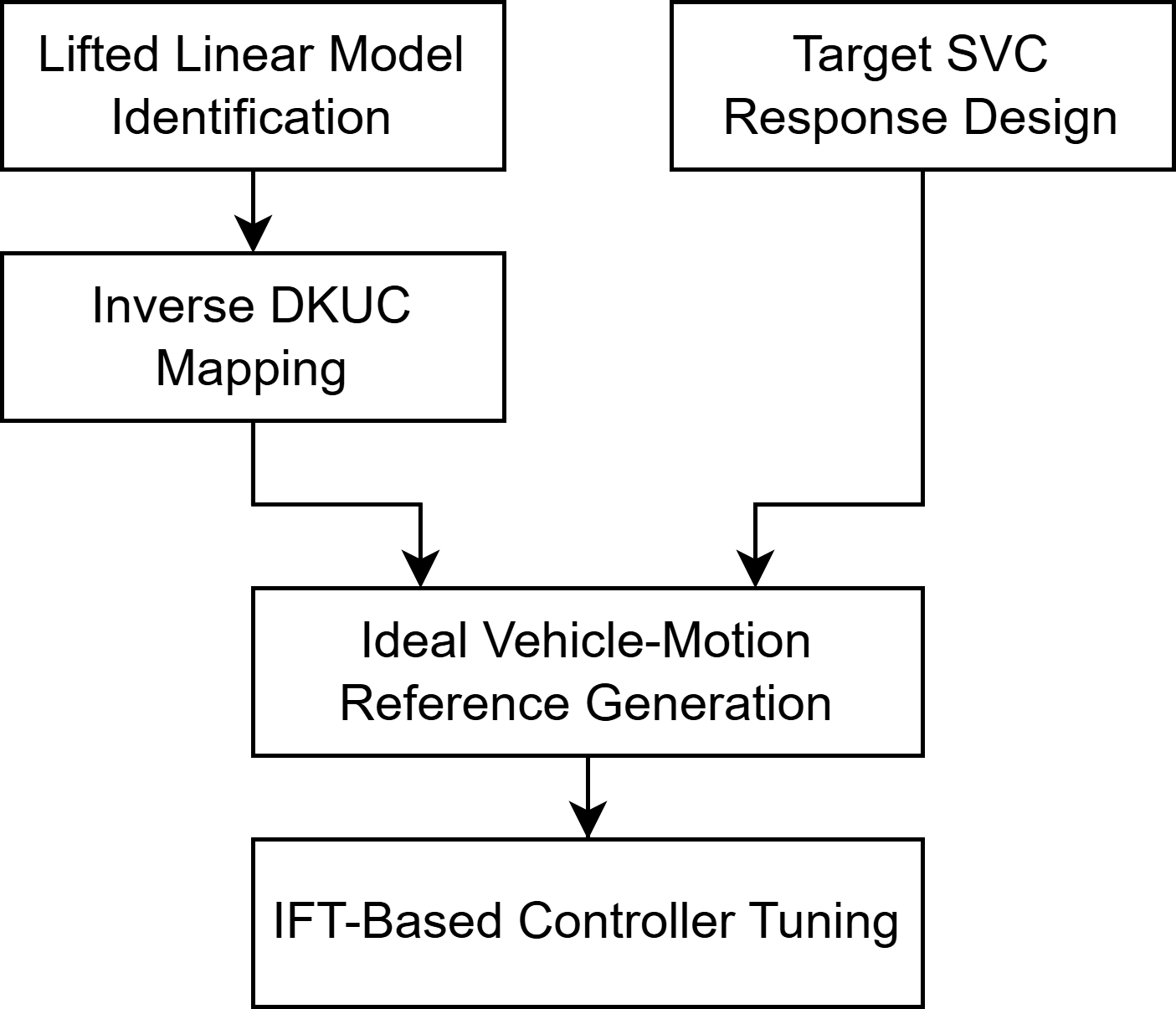}
        \caption{Design flow from a target SVC response to an ideal vehicle-motion response and fixed-structure controller tuning.}
        \label{fig:workflow}
        \vspace{-5.6mm}
    \end{figure}

    To construct this response, we use Koopman lifting, which approximates nonlinear dynamics by a linear predictor in an embedded space \cite{mauroy2020_springer,korda2018_automatica}.
    Deep KoopmanU with Control (DKUC) learns the embedded coordinates while retaining a linear input term \cite{shi2022_ral,xiao2023_transiv}.
    This input structure permits a least-squares inverse calculation: a prescribed SVC trajectory is mapped back to a vehicle-motion reference.
    The resulting vehicle-motion reference serves as the ideal response used for parameter tuning.
    Fig.~\ref{fig:workflow} summarizes these steps.

    The main contributions are as follows.
    \begin{enumerate}[label=(C\arabic*),leftmargin=2.7em,itemsep=0.2em]
        \item A DKUC-based inverse-design procedure constructs vehicle-motion references from SVC targets for tuning existing PI and LQR controllers.
        \item The reference design combines conflict-peak limiting, a target-magnitude floor, and inverse-response regularization; target selection compares requested SVC reduction with motion deviation.
        \item A numerical study quantifies prediction accuracy, target-selection tradeoffs, and reductions in regularized PI and LQR surrogate-tuning objectives.
    \end{enumerate}
    The proposed connection supplies the desired response needed to apply established tuning techniques to a physiological objective.

\section{Problem Setting and Design Objective}\label{sec:problem}
    \subsection{Vehicle--Passenger--6DoF-SVC System}
\begin{figure}[t]
            \centering
            \includegraphics[width=\linewidth]{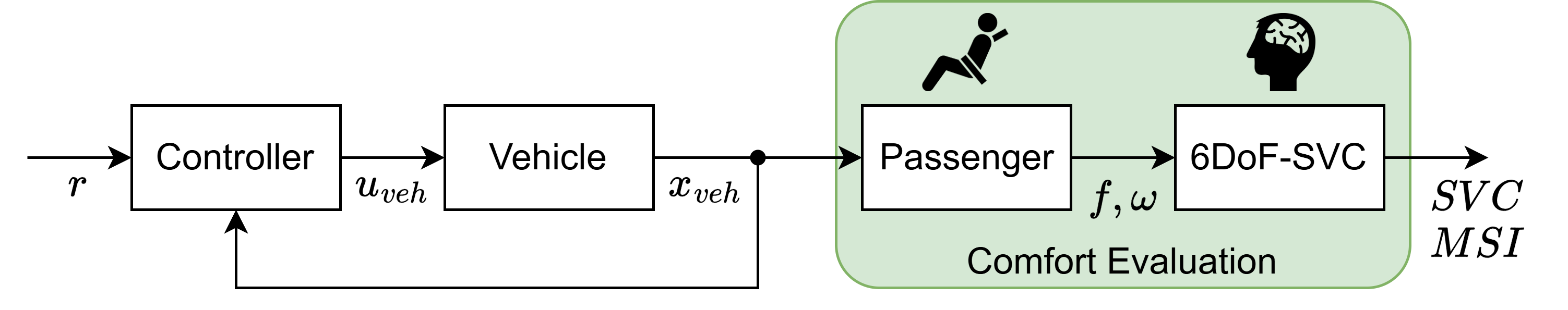}
            \caption{Vehicle control and comfort evaluation. Motor torques and steering produce vehicle motion, which drives the passenger--6DoF-SVC block.}
            \label{fig:overall_structure}
            \includegraphics[width=0.8\linewidth]{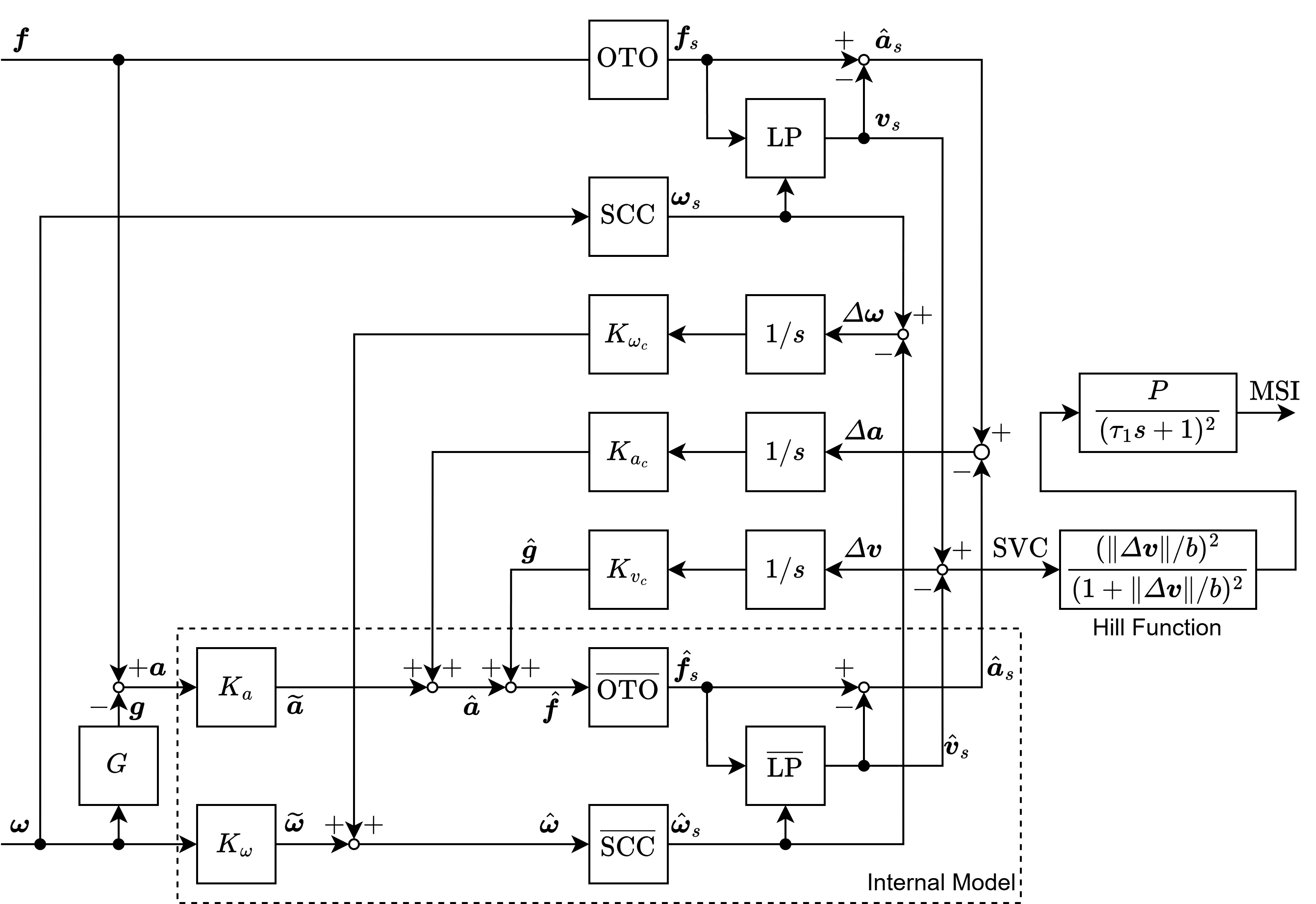}
            \caption{6DoF-SVC model used to evaluate sensory conflict from head motion. OTO, SCC, and LP denote otolith organs, semicircular canals, and a low-pass filter, respectively; overbars denote their internal models.}
            \label{fig:svc_model}
            \vspace{-3.5mm}
        \end{figure}

        In Fig.~\ref{fig:overall_structure}, the external command $r$ specifies the desired speed and lateral path.
        The vehicle controller follows this command through motor torques and steering, denoted by $u_{\mathrm{veh}}$ in the figure.
        The resulting vehicle response $x_{\mathrm{veh}}$ supplies motion to a separate passenger--6DoF-SVC evaluation block.

        The evaluation block's motion input and SVC output are denoted by
        \begin{align}
            u_{\mathrm{evo}}[k]&=
            \begin{bmatrix}
                \ddot p_x&\ddot p_y&\ddot p_z&\dot\theta_x&\dot\theta_y&\dot\theta_z
            \end{bmatrix}^{\top}\in\mathbb{R}^{6},
            \label{eq:uevo}\\
            x_{\mathrm{evo}}[k]&=
            \begin{bmatrix}\SVC_x&\SVC_y&\SVC_z\end{bmatrix}^{\top}\in\mathbb{R}^{3},
            \label{eq:xevo}
        \end{align}
        where $k$ is the sample index, $p_x,p_y,p_z$ are vehicle translations, and $\theta_x,\theta_y,\theta_z$ are roll, pitch, and yaw angles.
        The evaluation block thus maps vehicle accelerations and angular velocities to sensory conflict.
        The vehicle model used for data generation has no roll degree of freedom, so the six-channel interface has five active channels, with $\dot\theta_x[k]=0$.

        Within the evaluation block, the passenger is represented by the nonlinear two-stage inverted-pendulum model in \cite{makita2019_jsae}.
        Its roll and pitch dynamics are
        \begin{equation}
            \dot{x}_{\Phi}=f_{\mathrm{roll}}(x_{\Phi},u_{\Phi}),\qquad
            \dot{x}_{\Theta}=f_{\mathrm{pitch}}(x_{\Theta},u_{\Theta}),
        \end{equation}
        where $x_{\Phi}$ and $x_{\Theta}$ contain body and head angles and angular velocities, and $u_{\Phi}$ and $u_{\Theta}$ are accelerations at the seated position.
        This model supplies head-coordinate gravito-inertial acceleration $f\in\mathbb{R}^{3}$ and angular velocity $\omega\in\mathbb{R}^{3}$ to the 6DoF-SVC model in Fig.~\ref{fig:svc_model}.
        The present design uses its SVC components; motion sickness incidence (MSI) is a separate output of the physiological model.

    \subsection{Controller Structure}
        The design requirement is to retain the installed speed- and path-tracking laws and adjust the following parameters.
        The longitudinal PI controller generates the front and rear motor torques as
        \begin{equation}
            \begin{aligned}
                T_{m,i}(t)&=K_{P,i}e_{1,i}(t)+K_{I,i}\int_{0}^{t}e_{1,i}(\tau)\,d\tau, \\
                &\qquad i \in \{\mathrm{front},\mathrm{rear}\},
            \end{aligned}
            \label{eq:pi_control}
        \end{equation}
        where $e_{1,i}$ is the motor-speed tracking error.
        The tunable gains are collected in
        \begin{equation}
            \vartheta_{\mathrm{PI}}=
            \begin{bmatrix}
                K_{P,\mathrm{front}}&K_{I,\mathrm{front}}&K_{P,\mathrm{rear}}&K_{I,\mathrm{rear}}
            \end{bmatrix}^{\top}.
            \label{eq:theta_pi}
        \end{equation}

        The lateral LQR controller uses the state
        \begin{equation}
            x_{\eta}=
            \begin{bmatrix}
                \dot p_y&\dot\theta_z&p_y&\theta_z
            \end{bmatrix}
            ^{\top},
            \qquad \dot{x}_{\eta}=A_{\eta}(t)x_{\eta}+B_{\eta}\delta.
        \end{equation}
        The steering input $\delta$ is generated by
        \begin{equation}
            \delta=-K_Fx_{\eta}+Gx_I,
            \label{eq:lateral_control}
        \end{equation}
        where $x_I$ integrates the lateral-position error, $\dot{x}_I=p_{y,\mathrm{ref}}-p_y$.
        The gains $K_F$ and $G$ are obtained from the LQR design for the augmented state $\bar{x}_{\eta}=[x_{\eta}^{\top}\;x_I]^{\top}$.
        Tuning adjusts the state-weight entries $Q_{33}$ and $Q_{55}$ and the steering-input weight $R_{r2}$, which determine these gains:
        \begin{equation}
            \vartheta_{\mathrm{LQR}}=
            \begin{bmatrix}
                q_3&q_4&R_{r2}
            \end{bmatrix}
            ^{\top},
            \qquad q_3=Q_{33},\quad q_4=Q_{55}.
            \label{eq:theta_lqr}
        \end{equation}
        Here, $q_3$ weights lateral position and $q_4$ weights the integral state.
        A change in $\vartheta_{\mathrm{LQR}}$ changes the steering gains and hence the motion supplied to the evaluation block.
        In particular, the yaw-rate component $\dot\theta_z$ is shared by $x_{\eta}$ and $u_{\mathrm{evo}}$, while $\ddot p_y$ is the derivative of the first component of $x_{\eta}$, $\dot p_y$.

    \subsection{Design Objective}
        The command $r$ specifies the transport task; the physiological target $x_{\mathrm{ref}}$ specifies the desired SVC response.
        We construct $x_{\mathrm{ref}}$ from the baseline response by limiting conflict peaks while retaining a prescribed minimum target magnitude.
        The next step is to find a corresponding motion response $y_d$ in the same acceleration and angular-velocity coordinates as $u_{\mathrm{evo}}$:
        \begin{equation}
            x_{\mathrm{ref}}\xrightarrow{\;\text{inverse evaluation model}\;}y_d.
            \label{eq:physiology_to_response}
        \end{equation}
        The term \emph{ideal response} denotes this desired motion trajectory used as the tuning reference, without a guarantee of optimality or physical realizability.

        This intermediate response is needed because the evaluation-model input is motion, while the actuators accept motor torques and steering.
        The inverse model provides the motion to aim for; PI and LQR tuning is the separate step intended to realize it.
        The numerical implementation evaluates candidate parameters with reduced response models, called \emph{tuning surrogates}, as described in Section~\ref{sec:tuning}.

\section{DKUC Evaluation Model}
    \subsection{State-Inclusive Lifted Dynamics}
        We first identify the forward map from vehicle motion $u_{\mathrm{evo}}$ to sensory conflict $x_{\mathrm{evo}}$, which will then be inverted for reference generation.
        The dataset is $\mathcal{D}=\{(x_{\mathrm{evo}}[k],u_{\mathrm{evo}}[k])\}_{k=1}^{N}$.
        Following DKUC \cite{shi2022_ral}, a neural encoder $\psi_{\phi}$ adds learned observables to the measured SVC components:
        \begin{equation}
            z[k]=g_{\phi}(x_{\mathrm{evo}}[k])=
            \begin{bmatrix}
                x_{\mathrm{evo}}^{\top}[k]&\psi_{\phi}^{\top}(x_{\mathrm{evo}}[k])
            \end{bmatrix}^{\top}.
            \label{eq:lifting}
        \end{equation}
        With $z[k]\in\mathbb{R}^{n_z}$ and $C=[I_3\;0]$, the predictor is
        \begin{align}
            z[k+1]&=\mathcal{K}_Az[k]+\mathcal{K}_Bu_{\mathrm{evo}}[k],
            \label{eq:dkuc_pred}\\
            \hat{x}_{\mathrm{evo}}[k+1]&=Cz[k+1].
            \label{eq:dkuc_out}
        \end{align}
        Here, $\phi$ collects the encoder parameters, and $\mathcal{K}_A\in\mathbb{R}^{n_z\times n_z}$ and $\mathcal{K}_B\in\mathbb{R}^{n_z\times6}$ are the learned transition and input matrices.
        The SVC components are retained explicitly, and the additional observables are learned by the encoder without a prescribed dictionary.
        The matrix $C$ reconstructs SVC by linear selection, while $\mathcal{K}_B$ retains the motion input as a linear term for the inverse calculation.
        The encoder and training settings used in the numerical study are given in Table~\ref{tab:sim_setup}.

    \subsection{Training and Spectral Treatment}
        Training minimizes both SVC prediction error and the mismatch between recursively predicted and encoded lifted coordinates.
        Over a training window of $K$ steps, the implemented loss is
        \begin{align}
            L_{\mathrm{impl}}={}&\frac{w_{\mathrm{pred}}}{K}\sum_{k=1}^{K}\alpha_k
            \|\hat{x}_{k}-x_{k}\|_2^2
            \nonumber\\
            &+\frac{w_{\mathrm{lin}}}{K}\sum_{k=1}^{K}\alpha_k
            \|\hat{z}_{k}-g_{\phi}(x_k)\|_2^2
            \nonumber\\
            &+w_{\mathrm{stab}}L_{\mathrm{stab}}(\mathcal{K}_A),
            \label{eq:dkuc_loss}
        \end{align}
        where $x_k$ denotes the measured SVC vector within the window, $\hat{z}_k$ is the recursive prediction, and $\hat{x}_k=C\hat{z}_k$.
        The weights $\alpha_k=1+w_{\mathrm{low}}k/K$ increase the contribution of later prediction steps.

        The last term penalizes matrix properties that can amplify recursive predictions.
        Writing $K_{A,ij}$ for the entries of $\mathcal{K}_A$, define
        \begin{equation}
            d_i=\sum_{j\ne i}|K_{A,ij}|-\frac{1}{2}|K_{A,ii}|.
            \label{eq:diag_dominance}
        \end{equation}
        The implemented regularizer is
        \begin{align}
            L_{\mathrm{stab}}(\mathcal{K}_A)={}&
            \max\!\left(0,\frac{\|\mathcal{K}_A\|_F^2}{n_z}-c_F\right)
            \nonumber\\
            &+\max\!\left(0,\|\mathcal{K}_A\|_{\infty}-c_R\right)
            \nonumber\\
            &+\frac{1}{2n_z}\sum_{i=1}^{n_z}\max(0,d_i),
            \label{eq:stability_loss}
        \end{align}
        with $c_F=0.98$, $c_R=0.99$, and $\|\mathcal{K}_A\|_{\infty}=\max_i\sum_j|K_{A,ij}|$.
        The first two terms penalize matrix size, and the third promotes row diagonal dominance with a factor-of-two margin.
        This heuristic is complemented by a separate spectral projection that bounds the eigenvalues of the lifted transition.
        Every ten epochs and once after training, eigenvalues outside the disk of radius $\rho_{\mathrm{tar}}=0.995$ are rescaled by $0.999\rho_{\mathrm{tar}}/|\lambda_i|$.
        Six complex-conjugate pairs are also initialized in the $0.1$--$0.5~\mathrm{Hz}$ band.

    \subsection{Prediction and Modal Quantities}
        Forward prediction is assessed by propagating the lifted model with recorded motion inputs and comparing its output with the simulated SVC response.
        At horizon $k$, the mean Euclidean error is
        \begin{equation}
            E_k=\frac{1}{|\mathcal{T}_k|}\sum_{t\in\mathcal{T}_k}
            \|x_{\mathrm{evo}}[t+k]-\hat{x}_{\mathrm{evo}}[t+k\mid t]\|_2,
            \label{eq:mean_prediction_error}
        \end{equation}
        where $\mathcal{T}_k$ contains starting indices spaced by $20$ samples.
        To describe oscillation and decay in the lifted model, each complex eigenvalue $\lambda_i$ is associated with
        \begin{equation}
            f_i=\frac{|\arg(\lambda_i)|}{2\pi T_s},\qquad
            \zeta_i=-\frac{\operatorname{Re}(\ln\lambda_i)}{|\ln\lambda_i|},
            \label{eq:modal_metrics}
        \end{equation}
        where $T_s$ is the sampling period, $f_i$ is modal frequency, and $\zeta_i$ is damping ratio.
        Section~\ref{subsec:dkuc_prediction} uses the modal frequencies to check whether modes in the initialized $0.1$--$0.5~\mathrm{Hz}$ band remain after training.

\section{Ideal-Response Design}
    \subsection{Activity-Constrained SVC Target}
        The target is designed as a moderate modification of the baseline SVC response.
        Peak limiting reduces large conflict values while leaving smaller values initially unchanged.
        For each component $i\in\{x,y,z\}$ and threshold $\tau$, it gives
        \begin{equation}
            \tilde{x}_{\mathrm{ref},i}[k]=
            \begin{cases}
                x_{\mathrm{evo},i}[k], & |x_{\mathrm{evo},i}[k]|\le\tau,\\
                \tau\,\operatorname{sgn}(x_{\mathrm{evo},i}[k]), & |x_{\mathrm{evo},i}[k]|>\tau.
            \end{cases}
            \label{eq:svc_threshold}
        \end{equation}
        A magnitude floor is then imposed to avoid an almost inactive target sequence:
        \begin{equation}
            \|x_{\mathrm{ref}}[k]\|_2\ge
            \gamma\max_j\|\tilde{x}_{\mathrm{ref}}[j]\|_2,
            \qquad \gamma=0.1.
            \label{eq:activity_constraint}
        \end{equation}
        Here, activity refers to the magnitude of the SVC target.
        The floor regularizes this target before inverse mapping; vehicle-motion deviation is assessed after the mapping.

        The resulting target is characterized by its SVC reduction and the low-activity ratio
        \begin{equation}
            r_{\LA}=\frac{1}{N}\sum_{k=1}^{N}\mathbf{1}\!\left(
            \|x_{\mathrm{ref}}[k]\|_2<0.05\max_j\|x_{\mathrm{evo}}[j]\|_2\right),
            \label{eq:low_activity}
        \end{equation}
        where $N$ is the target-sequence length and $\mathbf{1}$ is the indicator function.
        Since a smaller target may require a larger change in vehicle motion, target selection also considers the deviation of the inverse-designed response from nominal motion.
        Section~\ref{sec:numerical} reports this target comparison and the resulting threshold selection.

    \subsection{Inverse DKUC Mapping}
        For a chosen target, $z_{\mathrm{ref}}[k]=g_{\phi}(x_{\mathrm{ref}}[k])$ specifies the desired lifted trajectory.
        The motion input must account for the part of the next desired state not supplied by the autonomous transition:
        \begin{equation}
            \Delta z_{\mathrm{ref}}[k]=z_{\mathrm{ref}}[k+1]-\mathcal{K}_Az_{\mathrm{ref}}[k].
            \label{eq:lifted_increment}
        \end{equation}
        Solving $\mathcal{K}_Bu_d[k]\approx\Delta z_{\mathrm{ref}}[k]$ gives the raw ideal motion response
        \begin{equation}
            u_d[k]=\mathcal{K}_B^{\dagger}
            \left(z_{\mathrm{ref}}[k+1]-\mathcal{K}_Az_{\mathrm{ref}}[k]\right),
            \label{eq:inverse_dkuc}
        \end{equation}
        where the pseudoinverse uses an absolute singular-value cutoff of $10^{-6}$.
        This is a minimum-norm least-squares solution over the retained singular directions.
        The desired lifted transition need not be reproduced exactly: with $P_B=\mathcal{K}_B\mathcal{K}_B^{\dagger}$, the residual is
        \begin{equation}
            r_z[k]=(I-P_B)\Delta z_{\mathrm{ref}}[k].
            \label{eq:inverse_residual}
        \end{equation}
        The residual measures the part of the desired transition that the computed input cannot reproduce.

        The raw sequence is then regularized to limit numerical excursions.
        It is first saturated at $\pm100$ in each channel's units ($\mathrm{m/s^2}$ for acceleration and $\mathrm{rad/s}$ for angular velocity).
        A channelwise $\kappa\sigma$ test with $\kappa=5$ and nominal-data standard deviation $\sigma$ replaces detected samples with the nominal sample at the same time index.
        Each channel is then clipped to $1.5$ times its nominal data range.
        The postprocessed sequence is denoted by $y_d$; its inactive roll-rate channel is zero.
        This nominal-data-based regularization produces the motion reference $y_d$ used in controller tuning.
        It is the ideal response of \eqref{eq:physiology_to_response} in the sense defined above: a design target, not a response verified to be realizable.

\section{IFT-Style Controller Tuning}\label{sec:tuning}
    With $y_d$ fixed, tuning seeks parameters whose motion response is close to this reference.
    The numerical implementation evaluates candidate parameters with the tuning surrogates introduced in Section~\ref{sec:problem}.
    For block $s\in\{\mathrm{PI},\mathrm{LQR}\}$, let $\tilde{y}_s[k;\vartheta_s]\in\mathbb{R}^{6}$ denote this tuning-surrogate response.
    The objective combines response mismatch and a parameter penalty:
    \begin{equation}
        J_s(\vartheta_s)=\frac{1}{N_t}\sum_{k=1}^{N_t}
        \|\tilde{y}_s[k;\vartheta_s]-y_d[k]\|_2^2
        +\lambda_s\|\vartheta_s\|_2^2,
        \label{eq:surrogate_cost}
    \end{equation}
    where $\vartheta_s$ is the corresponding parameter vector in \eqref{eq:theta_pi} or \eqref{eq:theta_lqr}, and $\lambda_{\mathrm{PI}}=10^{-3}$ and $\lambda_{\mathrm{LQR}}=10^{-4}$.
    The inverse in \eqref{eq:inverse_dkuc} uses successive target states, giving $N_t=4999$ motion samples from a $5000$-sample SVC target.
    Both objectives use all six motion channels without channel normalization.

    The PI surrogate is a gain-dependent first-order recursion; its integral gains affect only the parameter penalty in \eqref{eq:surrogate_cost}.
    The LQR surrogate uses the closed-loop spectral radius of a two-state bicycle model with sideslip and yaw-rate states.
    The same scalar entries $q_3$ and $q_4$ weight sideslip and yaw rate in the surrogate, whereas they weight lateral position and the integral state in the vehicle controller of \eqref{eq:theta_lqr}.
    These surrogates define the responses used in optimization; transfer of their tuned parameters to the vehicle controller requires separate validation.

    The two parameter blocks are optimized independently, in PI--LQR order, by sequential quadratic programming (SQP).
    The gradient is evaluated by central differences,
    \begin{equation}
        \frac{\partial J_s}{\partial\vartheta_{s,i}}
        \approx\frac{J_s(\vartheta_s+\varepsilon_i e_i)-J_s(\vartheta_s-\varepsilon_i e_i)}{2\varepsilon_i},
        \label{eq:central_difference}
    \end{equation}
    where $e_i$ is the $i$th coordinate vector and $\varepsilon_i$ is the parameter perturbation.
    This IFT-style procedure follows the desired-response tuning principle \cite{hjalmarsson1998_cs}, with gradients evaluated by finite differences of the surrogate costs.
    The maximum iteration count is $500$, and the optimality and step tolerances are $10^{-6}$.
    The parameter bounds are
    \begin{align}
        [0.1,0.01,0.1,0.01]^{\top}&\le\vartheta_{\mathrm{PI}}\le[20,10,20,10]^{\top},\\
        [0.01,0.01,0.01]^{\top}&\le\vartheta_{\mathrm{LQR}}\le[100,100,100]^{\top}.
    \end{align}

\section{Numerical Study}\label{sec:numerical}
    \subsection{Setup and Scope}
        The study evaluates prediction, ideal-response selection, and surrogate tuning in one driving scenario.
        The simulation uses MATLAB\textsuperscript{\textregistered}/Simulink\textsuperscript{\textregistered} R2025a, $T_s=0.01~\mathrm{s}$, and a duration of $100~\mathrm{s}$.
        The first $50~\mathrm{s}$ are used for DKUC training and the remainder for prediction testing.
        The latter interval is also used for target selection, controller tuning, and reporting; it is a holdout for DKUC identification, but the tuning results are in-sample.
        The longitudinal command switches among $60$, $50$, and $40~\mathrm{m/s}$ every $10~\mathrm{s}$.
        The lateral command is the sum of sinusoids with amplitudes $4$, $3$, and $5~\mathrm{m}$ at $0.50$, $1.25$, and $0.05~\mathrm{Hz}$, respectively.
        These are prescribed commands; the evaluation block is driven by the resulting vehicle motion $u_{\mathrm{evo}}$.
        Table~\ref{tab:sim_setup} lists the implementation settings.

        \begin{table}[t]
            \caption{Simulation and optimization settings.}
            \label{tab:sim_setup}
            \centering
            \small
            \begin{tabular}{@{}ll@{}}
                \toprule
                Item & Value\\
                \midrule
                Sampling period / duration & $0.01~\mathrm{s}$ / $100~\mathrm{s}$\\
                Data split & First half training, second half test\\
                Lifted dimension / horizon & $n_z=20$ / $K=100$\\
                Encoder hidden units & $128$, $64$ with $\tanh$ activation\\
                Learning rate / minibatch & $5\times10^{-4}$ / $2048$\\
                Training epochs & $2000$\\
                \multirow{2}{*}{Loss weights} & $(w_{\mathrm{pred}},w_{\mathrm{lin}},w_{\mathrm{low}},w_{\mathrm{stab}})$ \\
                &$=(5.0,0.1,2.0,0.15)$ \\
                Target spectral radius & $\rho_{\mathrm{tar}}=0.995$\\
                Tuning algorithm / horizon & SQP / $4999$ samples\\
                Maximum tuning iterations & $500$\\
                Optimality / step tolerance & $10^{-6}$ / $10^{-6}$\\
                \bottomrule
            \end{tabular}
        \end{table}

        The encoder has two fully connected hidden layers, and training uses Adam with gradient-norm clipping at $1.0$.
        The baseline LQR gains were computed at $V_x=1.0~\mathrm{m/s}$ and reused throughout the time-varying simulation, so their design operating point differs from the commanded speeds.

        These settings define an offline reference-generation and surrogate-tuning study under prescribed excitation.
        Full-model simulation data support DKUC identification and prediction assessment, while the reduced models in Section~\ref{sec:tuning} evaluate candidate controller parameters.
        Vehicle-level SVC and MSI improvements require a matched full-model comparison.

    \subsection{DKUC Prediction}\label{subsec:dkuc_prediction}
\begin{figure}[t]
            \centering
            \includegraphics[width=\columnwidth]{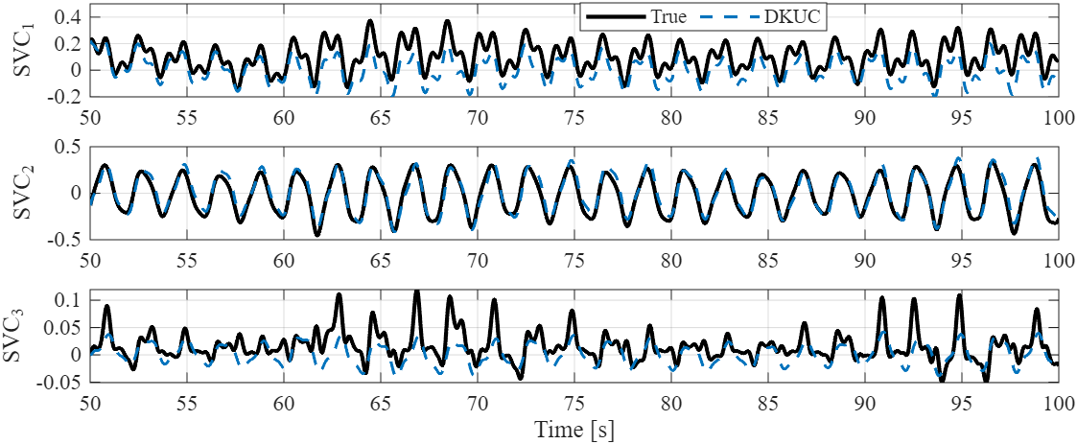}
            \caption{Simulated and recursively predicted SVC components over the test interval. Subscripts $1,2,3$ denote $x,y,z$, respectively.}
            \label{fig:dkuc_prediction}
            \vspace{-2mm}
        \end{figure}

        We first assess the forward predictor on which inverse reference generation relies.
        The errors in \eqref{eq:mean_prediction_error} are $E_{50}=5.41\times10^{-2}$, $E_{100}=6.00\times10^{-2}$, and $E_{200}=8.36\times10^{-2}$, corresponding to horizons of $0.5$, $1.0$, and $2.0~\mathrm{s}$.
        Fig.~\ref{fig:dkuc_prediction} shows recursive prediction over the test interval.
        The model captures the dominant oscillation of $\SVC_y$ and the principal trends of $\SVC_x$ and $\SVC_z$, while underestimating some sharp local peaks.

        The final spectral radius is $\rho(\mathcal{K}_A)=0.9940$, consistent with the imposed projection.
        Four eigenvalues, forming two conjugate pairs, remain in the initialized $0.1$--$0.5~\mathrm{Hz}$ band.

        \begin{figure*}[t]
            \centering
            \begin{subfigure}[b]{0.325\linewidth}
                \centering
                \includegraphics[width=\linewidth]{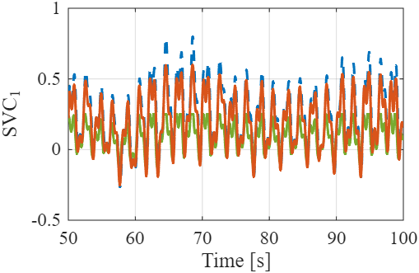}
                \caption{$\SVC_x$ response}
                \label{fig:opt_response_x1}
            \end{subfigure}
            \hfill
            \begin{subfigure}[b]{0.325\linewidth}
                \centering
                \includegraphics[width=\linewidth]{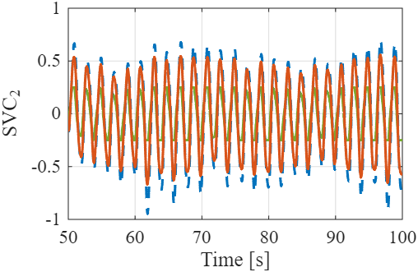}
                \caption{$\SVC_y$ response}
                \label{fig:opt_response_x2}
            \end{subfigure}
            \hfill
            \begin{subfigure}[b]{0.325\linewidth}
                \centering
                \includegraphics[width=\linewidth]{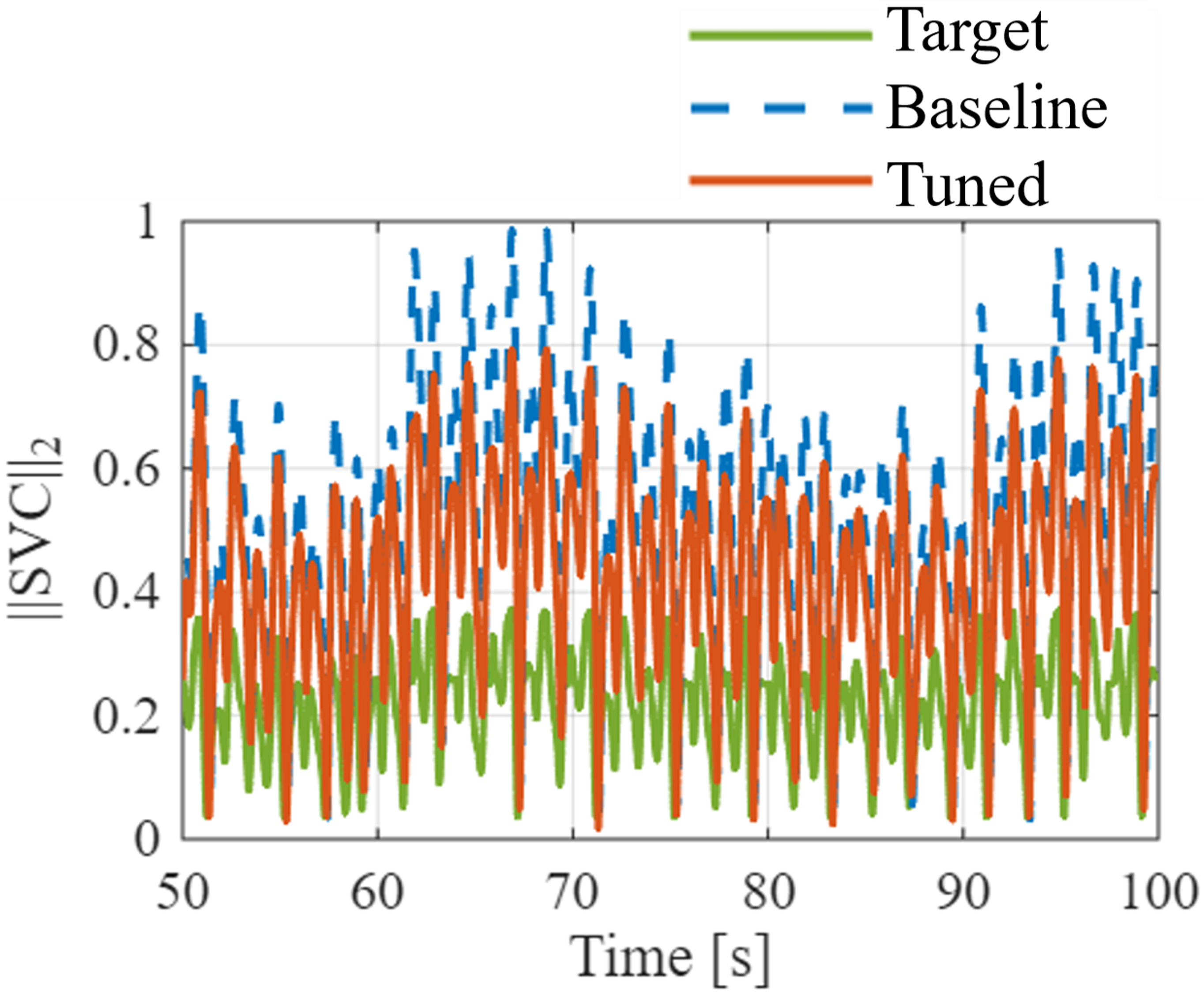}
                \caption{$\|\mathrm{SVC}\|_2$ response}
                \label{fig:opt_response_x3}
            \end{subfigure}
            \caption{Target and illustrative SVC-surrogate trajectories. The baseline and tuned cases are driven by $x_{\mathrm{evo}}$ and $x_{\mathrm{ref}}$, respectively. Axis labels $\SVC_1$ and $\SVC_2$ denote $\SVC_x$ and $\SVC_y$.}
            \label{fig:svc_results}
        \end{figure*}

    \subsection{Ideal-Response Selection}\label{subsec:target_selection}
        Table~\ref{tab:target_comparison} compares target conflict reduction with the motion change required by inverse design.
        Before activity adjustment, ratio candidates scale the baseline by $\alpha$, while the weighted $0.7/0.3$ candidate combines zero conflict and the baseline as $0.7\,\mathbf{0}+0.3x_{\mathrm{evo}}$.
        Threshold candidates use \eqref{eq:svc_threshold}.

        SVC reduction compares the norm of the entire target sequence $x_{\mathrm{ref}}$ with that of the baseline $x_{\mathrm{evo}}$; it is the requested reduction before controller tuning.
        Input deviation is the root-mean-square error (RMSE) between $y_d$ and nominal motion, pooled across unnormalized channels, and serves as a numerical comparison within this dataset.
        The low-activity ratio is defined by \eqref{eq:low_activity}.
        \footnote{Values within a relative tolerance of $10^{-12}$ of the low-activity threshold are treated as equality.}

        \begin{table}[t]
            \caption{Offline target selection: requested SVC reduction, deviation of the inverse-designed motion, and target low-activity ratio.}
            \label{tab:target_comparison}
            \centering
            \small
            \setlength{\tabcolsep}{3.2pt}
            \begin{tabular}{@{}lrrr@{}}
                \toprule
                Method & SVC red. & Input dev. & $r_{\LA}$\\
                \midrule
                Threshold $\tau=0.25$ & $6.95\%$ & $0.658$ & $0.00\%$\\
                Threshold $\tau=0.50$ & $-0.02\%$ & $0.531$ & $0.00\%$\\
                Ratio $\alpha=0.50$ & $49.99\%$ & $0.995$ & $0.00\%$\\
                Ratio $\alpha=0.70$ & $29.99\%$ & $0.765$ & $0.00\%$\\
                Weighted $0.7/0.3$ & $69.99\%$ & $1.239$ & $5.70\%$\\
                \bottomrule
            \end{tabular}
        \end{table}

        The ratio and weighted designs request larger conflict reductions but also produce larger motion deviations.
        The threshold $\tau=0.50$ leaves the target norm essentially unchanged ($-0.02\%$ reduction).
        A small negative reduction is possible because the magnitude floor in \eqref{eq:activity_constraint} can increase low-amplitude samples after peak clipping.
        We select $\tau=0.25$ as a conservative peak-limiting choice: it reduces the target norm by $6.95\%$, requires less motion deviation than the ratio designs, and has $r_{\LA}=0$.
        This offline comparison holds $\gamma=0.1$, $\kappa=5$, and the clipping factor $1.5$ fixed.

        The fitted input matrix has $\operatorname{cond}(\mathcal{K}_B)=5.80\times10^2$, used here as a numerical sensitivity diagnostic.
        The roll-rate input is unexcited in the identification data.

    \subsection{Surrogate Tuning Results}
        With the selected $y_d$ held fixed, the PI objective in \eqref{eq:surrogate_cost} decreases from $0.0869$ to $0.0439$, or $49.5\%$.
        The LQR objective decreases from $4.618$ to $3.126$, or $32.3\%$.
        These are reductions in the combined response-mismatch and parameter-penalty costs of the two tuning surrogates.

        Table~\ref{tab:surrogate_results} and Fig.~\ref{fig:svc_results} give an auxiliary illustration of SVC trajectories using a three-component recursion with output $\tilde{x}_{\SVC}$.
        Its parameter-dependent coefficients differ from those in the tuning models.
        It is driven by $x_{\mathrm{evo}}$ in the baseline case and by $x_{\mathrm{ref}}$ in the tuned case, rather than by $y_d$.
        The comparison illustrates the combined change in parameters and driving input; it does not isolate the contribution of controller tuning.

        The SVC comparison uses $N_e=5000$ samples, with mean absolute component $(3N_e)^{-1}\sum_{k=1}^{N_e}\|\tilde{x}_{\SVC}[k]\|_1$.
        Tracking RMSE and maximum absolute error are computed from componentwise errors relative to $x_{\mathrm{ref}}$.
        Cumulative SVC exposure is $T_s\sum_k\|\tilde{x}_{\SVC}[k]\|_2$, an accumulated surrogate-conflict measure distinct from MSI; the values are not percentages.

        \begin{table}[t]
            \caption{Illustrative SVC-surrogate comparison with changes in both parameters and driving inputs.}
            \label{tab:surrogate_results}
            \centering
            \small
            \begin{tabular}{@{}lrrr@{}}
                \toprule
                Metric & Baseline & Tuned & Change\\
                \midrule
                Tracking RMSE & $0.207$ & $0.151$ & $-27.0\%$\\
                Mean absolute component & $0.237$ & $0.203$ & $-14.1\%$\\
                Maximum absolute error & $0.697$ & $0.426$ & $-38.9\%$\\
                Cumulative SVC exposure & $25.72$ & $22.03$ & $-14.3\%$\\
                \bottomrule
            \end{tabular}
            \vspace{-3mm}
        \end{table}

\section{Discussion}
    \subsection{Role of Ideal-Response Design}
        The motion reference $y_d$ provides a common interface between physiological target selection and fixed-structure controller tuning.
        The target comparison makes the tradeoff between requested conflict reduction and motion deviation explicit, and the surrogate optimization demonstrates use of the selected reference in parameter tuning.

    \subsection{Inverse Realizability and Numerical Sensitivity}
        Physical realization of $y_d$ requires responses that satisfy vehicle-state and actuator constraints.
        The lifted residual in \eqref{eq:inverse_residual} remains to be evaluated alongside full-model tracking of $y_d$ to assess inverse consistency and realizability.

    \subsection{Target Selection and Validation Requirements}
        Full-model validation should apply the tuned PI and augmented LQR controllers under matched commands, initial conditions, passenger parameters, and simulation horizons, evaluating head motion through the original 6DoF-SVC model.
        Further scenarios and sensitivity analysis of $\tau$, $\gamma$, $\kappa$, and clipping will guide target selection across operating conditions.
        Because the encoder currently uses only the instantaneous SVC components, adding past SVC and motion samples is a possible extension for representing unobserved evaluation-model memory.

\section{Conclusion}
    This paper presented a procedure for constructing the desired vehicle-motion response needed in motion-sickness-aware tuning of an existing controller.
    A peak-limited, activity-constrained SVC target is mapped to a regularized motion reference through an inverse DKUC model.
    The reference is then supplied to fixed-structure PI and LQR surrogate tuning.
    For the considered data, the selected target has a $6.95\%$ lower SVC-sequence norm than the baseline, and tuning reduces the regularized PI and LQR surrogate objectives by $49.5\%$ and $32.3\%$.
    The resulting interface makes the physiological objective explicit in the response used for parameter tuning while retaining the controller structures.

\section*{Acknowledgment}
    The authors used the generative AI tools Claude Opus 5 (Anthropic) and GPT-5.6 Sol (OpenAI) for language editing, reviewed the text, and retain full responsibility for its content.

\bibliographystyle{IEEEtran}
\bibliography{reference}

\end{document}